\documentclass[prl,aps,twocolumn,preprintnumbers,showpacs]{revtex4}
\pdfoutput=1

\usepackage{amsmath,amssymb,bm}
\usepackage{graphicx}
\usepackage{color}

\begin{document}
\title{Twist instability in strongly correlated carbon nanotubes }
\author{Wei Chen}
\affiliation{Department of Physics, University of Washington, Seattle, Washington
98195-1560, USA}

\author{A. V. Andreev}
\affiliation{Department of Physics, University of Washington, Seattle, Washington
98195-1560, USA}
\author{A.  M. Tsvelik}
\affiliation{Department of  Condensed Matter Physics and Materials Science,
Brookhaven National Laboratory, Upton, NY 11973-5000, USA}
\author{Dror Orgad}
\affiliation{Racah Institute of Physics, The Hebrew University, Jerusalem 91904,
Israel}

\date{July 14, 2008}
\begin{abstract}
We show that strong Luttinger correlations of the electron liquid in
armchair carbon nanotubes lead to a significant enhancement of the
onset temperature of the putative twist Peierls instability. The
instability results in a spontaneous uniform twist deformation of
the lattice at low temperatures, and a gapped ground state.
Depending on values of the coupling constants the umklapp electron
scattering processes can assist or compete with the twist
instability. In case of the competition the umklapp processes win in
wide tubes. In narrow tubes the outcome of the competition depends
on the relative strength of the e-e and e-ph backscattering. Our
estimates show that the twist instability may be realized in free
standing $(5,5)$ tubes.
\end{abstract}
\maketitle

The band structure of armchair carbon nanotubes is metallic~\cite{Saito}.
However, due to the Peierls mechanism the metallic state may become unstable at
low temperatures. For undoped armchair tubes the possible instability modes are a
lattice deformation with a finite wave vector~\cite{Mintmire1992} and the twist
instability at zero wave vector~\cite{Figge,Suzuura2000}. Previous studies of
this instability were carried out either in the noninteracting electron
approximation~\cite{Mintmire1992,Viet,Ando,Figge,Suzuura2000}, yielding a very
low instability temperature $\lesssim 0.1 K$ even for the thin $(5,5)$ nanotubes,
or using \emph{ab initio} calculations in the local density
approximation~\cite{LDA}. These treatments can not account for the Luttinger
liquid effects, which are expected to be pronounced in these
systems~\cite{Egger1997,Kane1997}.

Here we study the possibility of the Peierls instability in armchair carbon
nanotubes in the presence of electron-electron (e-e) interactions. We find that
in the absence of backscattering processes the Luttinger liquid correlation of
electrons significantly enhance the onset temperature for the \emph{twist}
Peierls instability. The enhanced transition temperature in narrow tubes can be
easily reached experimentally. We show that depending on the values of the e-e
umklapp coupling constants the umklapp
processes~\cite{Egger1997,Kane1997,Odintsov1999,Nersesyan2003} either enhance or
compete with the twist instability. The energy gap due to the umklapp processes
increases slower with decreasing tube radius $R$ than the gap due to the Peierls
instability. In case of the competition the twist instability can occur only in
narrow tubes. Based on the theoretical values~\cite{Jishi1993,Mahan} of the
electron-phonon  couplings we estimate the Peierls gap in $(5,5)$ tubes with
$R\sim 0.35 \, nm$ to be of the order of $\sim70\, K$. The gaps due to the
umklapp processes are much harder to evaluate theoretically because of their
strong sensitivity to the short range part of the Coulomb repulsion
potential~\cite{Odintsov1999,Nersesyan2003}. Gate voltage scans of conductance
for armchair tubes~\cite{Ouyang2001,Dai2003} indicate that the umklapp gaps for
$(5,5)$ tubes are of order or below our estimates for the twist gap. This
suggests that the Peierls instability may be realized in torsional nanomechanical
resonators~\cite{Washburn} based on $(5,5)$ armchair tubes.

\begin{figure}[ptb]
\includegraphics[width=8.0cm]{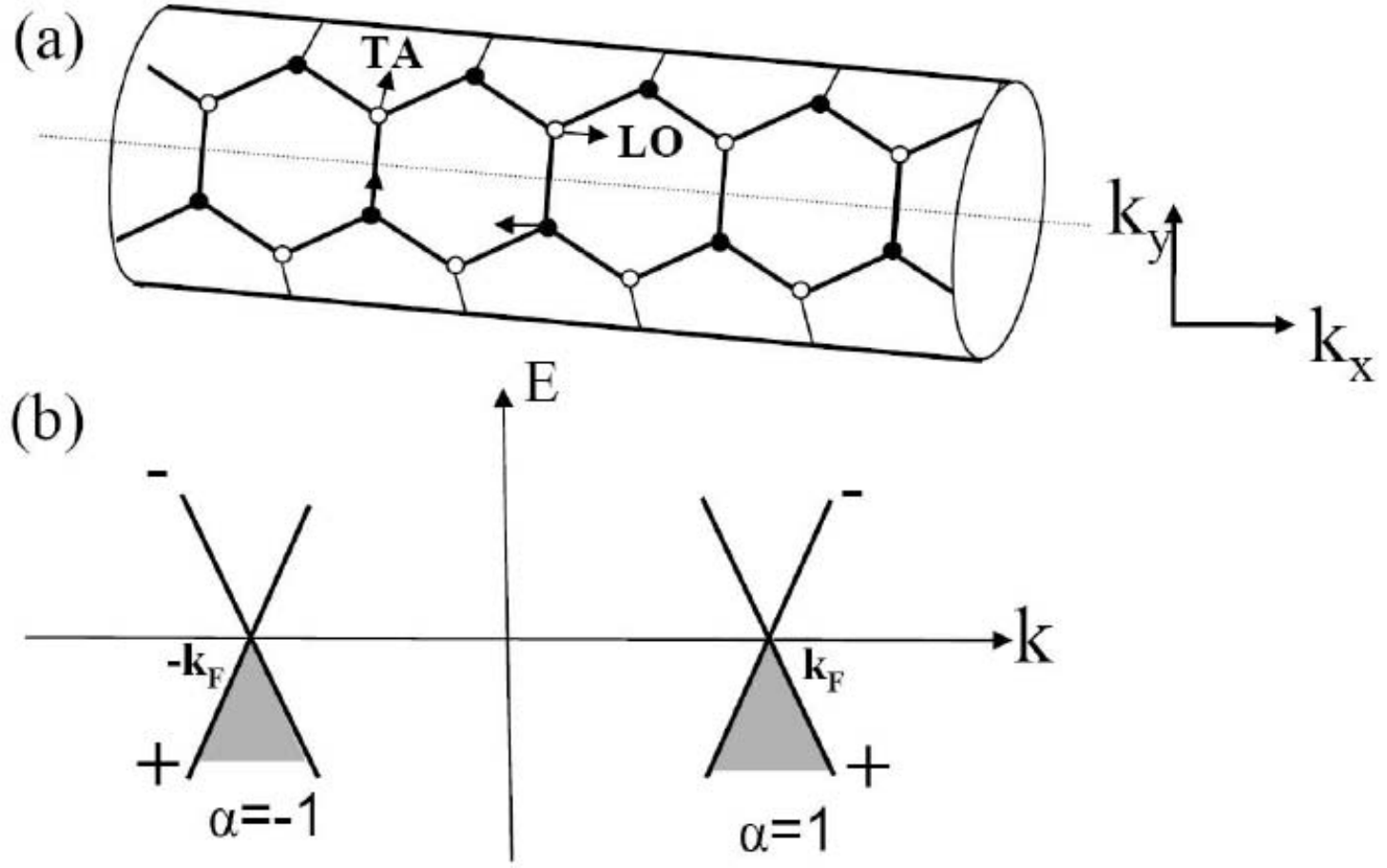}
\caption{(a) Schematic picture of an armchair tube. Filled and open
circles denote atoms in the A and B sublattices. The arrows show
atomic displacements in the TA and LO phonon modes. (b) Free
electron spectrum of armchair carbon nanotubes: $+$ and $-$ denote
the parity of the bands. $\alpha= \pm 1$ denotes the two valleys
with the Fermi points at $\pm k_{F}$.} \label{fig:1}
\end{figure}

We consider an undoped free standing $(N,N)$ armchair carbon
nanotube, see Fig.~\ref{fig:1} (a).  It is symmetric with respect to
reflection in the plane that goes through the tube axis and maps the
A and B sublattices onto each other. Thus both electron and phonon
modes are characterized by parity, $\pm 1$~\cite{Figge}. The low
energy electron spectrum is formed by two bands with opposite parity
and zero angular momentum along the tube axis. These bands intersect
at the two Dirac points as shown in Fig.~\ref{fig:1} (b).

Only the phonon modes with zero angular momentum along the tube axis couple to
the low energy electron modes. Furthermore,  only the backscattering part of the
electron-phonon (e-ph) coupling can lead to the Peierls instability and is
strongly enhanced by the Luttinger liquid correlations.  We therefore neglect the
forward scattering part of the e-ph coupling~\footnote{For large values of the
coupling constant the forward scattering part of the e-ph coupling can become
important and lead to the Bardeen-Wentzel instability~\cite{Martino2003}.
However, for values of the e-ph coupling quoted in the
literature~\cite{Jishi1993} this does not occur even in the most narrow tubes,
and accounting for this interaction leads only to inessential corrections to our
treatment.}. The backscattering by positive parity phonons involves large
momentum transfer and corresponds to phonons with rather high frequencies, $\sim
1000 K$. This cuts off Luttinger liquid renormalization and results only in a
finite softening of these phonons~\cite{softening}. The Peierls instability
arises from the coupling to the negative parity phonons with small momentum.
There are three such phonon modes: transverse acoustic (TA) and
longitudunal/radial optical (LO/RO)~\cite{Figge, Mahan}. The e-ph coupling arises
from the change in the hopping matrix element due to the atomic displacements.
The coupling to the RO mode is due to the curvature of the tube and is smaller
than the coupling to the LO and TA modes in $1/N$~\cite{Mahan}. We neglect it
below. Its inclusion would only enhance the twist instability.

The phonon Hamiltonian can be written as
\begin{equation}\label{eq:H_0ph}
H_{ph}= \frac{\rho}{2}  \sum_{q} \sum_{a=L,T}  (|\dot{u}_a(q)|^2
+\omega_a(q)^2|u_a(q)|^2).
\end{equation}
Here  $\rho$ is the mass per unit length of the tube and $u_a(q)$
are Fourier components of atomic displacements. The index $a=L$
corresponds to the LO mode with momentum independent frequency
$\omega_L(q)=\omega_o$ and $a=T$ corresponds to the TA mode with
frequency $\omega_T(q)=s_T q$, $s_T$ being the speed of sound.

The free electron Hamiltonian can be written as
\begin{equation}\label{eq:H_0e}
H_{0e}=-i\hbar \, v_{F}\sum_{\alpha r \sigma}\int dx\, \psi_{\alpha r
\sigma}^{+}(x) \, r\,\partial_{x}\psi_{\alpha r \sigma}(x).
\end{equation}
Here $v_F \approx 8 \times 10^5\, m/s$ is the Fermi velocity, $\alpha=\pm 1$ is
the valley index, $r=\pm1$ represents right and left movers, and $\sigma$ is the
electron spin.

The forward scattering part of the e-e interaction is much stronger than the
backscattering part, the latter being small in $1/N$~\cite{Balents1997}. The
former can be written as
\begin{equation}
H_{\rho}=g_{\rho}\int dx\,  n^2 (x), \quad n(x)=\sum_{\alpha r
\sigma} \psi^{+}_{\alpha r \sigma}(x) \psi_{\alpha r \sigma}(x).
\end{equation}
Here we assumed that the Coulomb interaction is screened by e.g. a
gate at a distance $d$, and the coupling constant being $g_{\rho}
\approx2 e^2 \ln (d/R)$.

The above electron Hamiltonian is bosonized by the standard
procedure (see, for example, ~\cite{Giamarchi, TsvelikBook})
\begin{equation}\label{eq:bosonization}
\psi_{\alpha r \sigma}(x)=\frac{F_{\alpha \sigma}}{\sqrt{2\pi \xi}}
\exp\{-i\sqrt{\pi}[\Theta_{\alpha \sigma}(x)-r\, \Phi_{\alpha
\sigma}(x)]\}.
\end{equation}
Here $F_{\alpha \sigma}$ are the Klein factors, $\xi$ is a short
distance of order of the radius of the nanotube, and $\Phi$ and
$\Theta$ are boson fields satisfying the commutation relation
\begin{equation}
[\Phi_{\alpha \sigma}(x),\Theta_{\alpha^{\prime}
\sigma^{\prime}}(x^{\prime})]=-i \delta_{\alpha \alpha^{\prime}}
\delta_{\sigma \sigma^{\prime}} \theta(x-x^{\prime}),
\end{equation}
with $\theta(x-x^{\prime})$ being the step function. Introducing charge and spin
modes combining different valleys,
\begin{eqnarray}
\Phi_{\alpha \sigma}&=& \left[ \Phi_{c +} + \alpha \Phi_{c -} + \sigma
\Phi_{s +} +\alpha \sigma \Phi_{s -} \right]/2, \nonumber \\
\Theta_{\alpha \sigma }&=& \left[ \Theta_{c +} + \alpha \Theta_{c -} + \sigma
\Theta_{s +} +\alpha \sigma \Theta_{s -} \right]/2, \nonumber
\end{eqnarray}
we rewrite the forward scattering part of the electron Hamiltonian as
\begin{equation}\label{eq:H_forward}
H_{0e}+ H_{\rho}= \sum_{j}\frac{\hbar u_{j}}{2}\int dx
{[K^{-1}_{j}(\nabla \Phi_{j})^2+K_{j}(\nabla \Theta_{j})^2]}.
\end{equation}
Here $j=c\pm, s\pm$, $u_{j}=v_{F}/K_{j}$ with $K_j$ being the
Luttinger parameter is the velocity of the $j$-th mode. For the
three modes $j=c-, s\pm$, $K_j=1$. For the charge mode
$K_{c+}=1/\sqrt{1+4 g_{\rho}/\pi v_{F}\hbar}$. Below we will use the
high charge stiffness approximation, $K_{c+}\ll 1$.

The coupling of electrons to the TA and LO phonons is described by the
Hamiltonian
\begin{equation}
H_{ep}=  \int dx M(x)  [ g_T \nabla u_{T}(x)+g_L u_{L}(x)],
\end{equation}
with $M(x)=-i \sum_{\alpha r \sigma}\alpha r \psi^{\dagger}_{\alpha r \sigma}(x)
\psi_{\alpha -r \sigma}$. The bosonized form of the operator $M$ is
 \begin{eqnarray}\label{eq:H_ep}
M&=& -\frac{4}{\pi
 \xi}\left[\prod_{\nu=\pm}\cos(\sqrt{\pi}\Phi_{c\nu})
 \sin(\sqrt{\pi}\Phi_{s\nu})
 \right.\nonumber \\
 &&\left. +\prod_{\nu=\pm}\sin(\sqrt{\pi}\Phi_{c\nu})
 \cos(\sqrt{\pi}\Phi_{s\nu}) \right].
\end{eqnarray}
By writing $M$ in this form we adopted the convention
$F^{\dag}_{\alpha\sigma}F_{\alpha\sigma}=1$. For $K_{c+} \ll 1$ the e-ph coupling
(\ref{eq:H_ep}) leads to a strong renormalization of the phonon mode. To the
second order in the e-ph coupling the phonon propagator matrix becomes
$\mathcal{D}(\omega_n,q)=[\mathcal{D}_0^{-1}(\omega_n,q)
+\Sigma(\omega_n,q)]^{-1}$, where $\omega_n=2n\pi T$ is the Matsubara frequency,
$\mathcal{D}_0^{-1}(\omega_n,q)=\delta_{a,a'} \rho[\omega^2_n +\omega^2_a(q)]$ is
the bare phonon propagator, and $\Sigma(\omega_n,q)$ is the self energy due to
the e-ph interaction. The latter has the form
\begin{equation}
\Sigma(\omega_n,q)=- P( \omega_n, q)\left(
  \begin{array}{cc}
    g_{T}^2 q^2 &  -i g_{T} g_{L}q \\
    i g_{L} g_{T}q & g^2_{L} \\
  \end{array}
\right).
\end{equation}
where $P(\omega_n,q) =  \int d r e^{i\omega_n \tau -iq x} \langle
M(0)M(r)\rangle$ and $\langle \ldots \rangle$ denotes thermal
averaging with respect to the forward scattering Hamiltonian
(\ref{eq:H_forward}). In the bosonized representation it is given by
\begin{eqnarray}
P(\omega_n,q) &=&  \int  \frac{2 d^2 r}{(\pi \xi)^2} \, e^{i\omega_n
\tau -iq x-\frac{\pi}{2}\sum_j \langle
[\Phi_{j}(r)-\Phi_{j}(0)]^{2}\rangle }. \nonumber
\end{eqnarray}
In the long wavelength limit, $q, \omega \to 0$ and for $K_{c+} \ll 1$ we obtain,
\begin{equation}
P(0,0)\equiv P =\frac{1}{v_{F}}\left(\frac{\beta v_{F}}{\pi
\xi}\right)^{\frac{1}{2}}\frac{1}{\pi^{2}}B^{2} \left(3/8,1/4\right),
\end{equation}
where $B(a,b)$ is the Euler Beta function.

The renormalized phonon frequencies are given by the poles of
$\det{(\mathcal{D}(\omega_n,q))}$ analytically continued to real frequencies. The
instability first appears when the renormalized frequency of the acoustic mode,
$\tilde\omega_T$ turns to zero,
\begin{equation}
\tilde \omega^2_T \equiv \omega^2_T \, \frac{1-(\tilde g^2_T +\tilde g^2_L)v_F
P}{1-\tilde g^2_L v_F P}=0,
\end{equation}
where $\tilde g_T = g_T/\sqrt{\rho s^2_T v_F}$ and $\tilde g_L=g_L/\sqrt{\rho
\omega^2_o v_F}$ are the characteristic dimensionless e-ph coupling constants for
TA and LO phonons respectively. The  mean field twist instability temperature is
then
\begin{equation}\label{eq:Tc}
T_{c}=(\tilde g^2_T+\tilde g^2_L)^2 \frac{v_F}{\pi^5
\xi}B^{4}\left(\frac{3}{8},\frac{1}{4}\right).
\end{equation}
In one dimension fluctuations shift mean field instabilities to zero
temperature. In the given case the order parameter $ \nabla u$ is
real and the instability is of the Ising type. As is the case for
the Ising model, a finite order parameter exists only at $T=0$
whereas at finite $T$ we have a state with a finite density of
solitons (domain walls separating areas with different sign of
$\nabla u$). The mean field $T_c$ is, however, not devoid of meaning
and gives an estimate of the spectral gaps associated with the
strong electron-phonon coupling. The exact composition of the
excitation spectrum is currently unknown. Using  the Feynmann's
variational principle we can estimate the magnitudes of the gaps of
the singlet excitations of fields $\Phi_j$~\cite{Giamarchi}. These
excitations represent small fluctuations around the minima of the
action. In addition to them there are kinks of these fields. More
complete analysis of the spectrum will be given in the extended
version of this paper. For the condensed state, we assume the
lattice has a static deformation, $\nabla u_T=\eta$, $u_L=\xi$,  and
for the electronic degrees of freedom we use a variational action of
the form,
\begin{equation}\label{eq:S_trial}
S_0= \frac{1}{2}\sum_j \int \frac{dxd\tau}{ K_j u_j}\left[\dot
\Phi_j^2+u_j^2 (\nabla \Phi)^2+ \Delta^2_j\Phi^2_j\right],
\end{equation}
where $\Delta_j$  are variational parameters. The variational free
energy for electrons is $F =F_0 + T\langle S-S_0\rangle_0$, where
$\langle \ldots \rangle_0$ denotes averaging with respect to the
trial action (\ref{eq:S_trial}). Minimizing the total free energy of
both the lattice and electrons with respect to $\Delta_j$ and the
static lattice deformations $\zeta$ and $\eta$ one obtains the
following equations in the $K_{c+}\to 0$ limit,
\begin{eqnarray}
\eta &=&\frac{1}{\pi \xi}\frac{4g_T}{\rho s^2_T}\prod_{j\neq
c+}\left(\frac{\Delta_j}{v_F \Lambda}\right)^{\frac{1}{4}}, \\
\zeta&=&\frac{1}{\pi \xi}\frac{4g_L}{\rho \omega^2_o}\prod_{j\neq
c+}\left(\frac{\Delta_j}{v_F \Lambda}\right)^{\frac{1}{4}}, \\
 \frac{\Delta^2_j}{2 K_j u_j}&=&\frac{1}{
\xi}(g_T \eta+ g_L \zeta)\prod_{j\neq c+}\left (\frac{\Delta_j}{v_F
\Lambda}\right)^{\frac{1}{4}}.
\end{eqnarray}
 This gives identical gaps for the four electron modes,
\begin{equation}\label{eq:gap}
\Delta_j=\frac{v_F}{\xi}\left(\frac{8(\tilde g^2_T+\tilde
g^2_L)}{\pi }\right)^2,
\end{equation}
and the spontaneous twist angle
\begin{equation}
\eta= \frac{\tilde g_T}{\sqrt{\frac{1}{2\pi}(\tilde g^2_T+\tilde
g^2_L)}}\frac{\Delta}{\sqrt{\rho s^2_T v_F}}.
\end{equation}

In the framework of the tight binding model, the values of the coupling constants
can be expressed~\cite{Barisic1970,Frohlich1952,Jishi1993,Mahan} in terms of the
derivative, $\frac{\partial J(r)}{\partial r}$, of the transfer integral $J(r)$
with respect to the bond length $r$. For armchair nanotubes, these coupling
constants were obtained to be: $g_{T}=\frac{\sqrt{3}}{4} a \frac{\partial
J(r)}{\partial r}$ and $g_{L}=\frac{3}{2}\frac{\partial J(r)}{\partial r}$, where
$a=2.5\, {\AA}$ is the graphene lattice constant and $\frac{\partial
J(r)}{\partial r}=-\lambda J_{0}/c$ with $J_{0}=2.6\, eV$ the hopping integral,
and $c=1.4 \,{\AA}$ the bond length ~\cite{Smalley2001}. Here $\lambda$ is a
dimensionless constant, whose theoretical value is 2~\cite{Ando2002}. The linear
mass density for an $(N,N)$ armchair nanotube is $\rho=4NM/a$ with $M$ being the
carbon atom mass. The twiston phonon velocity $s_T \sim 1.4\times 10^4 m/s$ and
the LO phonon energy $\omega_0\sim0.18\, eV$~\cite{Figge}. Thus both
$\tilde{g}_T$ and $\tilde{g}_L$ are proportional to $1/\sqrt{R}$. For a $(5,5)$
armchair nanotube with  $R\sim 0.35 \, nm$, the transition temperature from
Eq.~(\ref{eq:Tc}) is about $40\, K$ and the gap from Eq.~(\ref{eq:gap}) is about
$70 \,K$. As expected, the mean field instability temperature and the gap have
the same order of magnitude. The twist angle is $\eta \sim 3\times10^{-4}$.

As mentioned above the soliton effects will restore the symmetry at
finite temperatures. However, if the length of the tube is shorter
than the typical intersoliton distance, at experimental time scales
the system will appear twisted in one direction. The estimated
characteristic twist angle for $(5,5)$ nanotubes, $\sim 0.01^\circ$,
is too small to be detected by STM imaging. However, for a freely
suspended tube clamped at one end the accumulated rotation angle at
a distance $x$ from the clamp, $\theta(x)=\eta x/R$, becomes
substantial for $ x \sim 1 \, \mu m$.  This twist can be detected in
torsional nanomechanical resonators similar to the those studied in
Ref.~\onlinecite{Washburn} by measuring a deflection angle of a
paddle attached  to an armchair tube.

The discussion above ignored the backscattering part of the e-e interactions. The
most relevant backscattering terms correspond to the so-called umklapp processes,
which transfer two right-moving electrons into left moving ones or vice
versa~\cite{Kane1997,Egger1997,Odintsov1999,Nersesyan2003}. The bosonized form of
the umklapp interaction can be written as~\cite{Nersesyan2003}
\begin{eqnarray}\label{eq:H_Um}
                {H}_{u} &=&
-\frac{1}{2(\pi \xi)^2}\int dx \cos(\sqrt{4\pi}\Phi_{c+})\nonumber\\
                &&\left\{g_3 \cos(\sqrt{4\pi}\Theta_{s-})
                + (g_3-g_1)\cos(\sqrt{4\pi}\Phi_{s+})
                \right.\nonumber\\
                &&\left.+ g_1[\cos(\sqrt{4\pi}\Phi_{c-})-
                \cos(\sqrt{4\pi}\Phi_{s-})]
\right\},
\end{eqnarray}
where the coupling constants can be expressed in terms of the matrix
elements of the Coulomb interaction of electrons on the two
sublattices with a large ($q \sim 2k_F$) and small ($q\sim 0$)
momentum transfer: $g_3=V_{AA}(2k_F)+V_{AB}(2k_F)$ and
$g_1=V_{AA}(0)-V_{AB}(0)$~\cite{Odintsov1999,Nersesyan2003} . In the
absence of electron phonon coupling the umklapp processes lead to an
insulating state with four different gaps for the different
modes~\cite{Nersesyan2003}. The properties of the ground state and
the magnitudes of the gaps depend on the values of the coupling
constants $g_1$ and $g_3$. These constants are very difficult to
evaluate because they depend very sensitively on the short range
part of the Coulomb interaction and on the spatial dependence of the
electron density in the atomic orbitals. Within the model of
point-like atomic orbitals and Coulomb interaction with a short
distance cutoff numerical evaluation~\cite{Odintsov1999} of the
matrix elements gives $g_3> g_1>0$. In this case it is easy to see
from Eqs.~(\ref{eq:H_Um}) and (\ref{eq:H_ep}) that the umklapp
processes compete with the twist instability. Indeed the umklapp
interaction favors the condensation of the fields $(\Phi_{c+},
\Phi_{c-}, \Phi_{s+}, \Theta_{s-})$ in the ground state to
$(0,0,0,0)$, while the twist instability results in condensation of
fields $(\Phi_{c+}, \Phi_{c-}, \Phi_{s+}, \Phi_{s-})$ to $(0,0,
\frac{\sqrt{\pi}}{2}, \frac{\sqrt{\pi}}{2})$ or
$(\frac{\sqrt{\pi}}{2}, \frac{\sqrt{\pi}}{2}, 0,0)$. We note that
for $g_1>g_3>0$ the umklapp processes enhance the twist instability
and favor the ground state condensation pattern $(\Phi_{c+},
\Phi_{c-}, \Phi_{s+}, \Phi_{s-})=   (0,0, \frac{\sqrt{\pi}}{2},
\frac{\sqrt{\pi}}{2})$.

For $g_3>g_1>0$ the fate of the low temperature state thus depends on the
relative strength of the e-ph and umklapp coupling constants. To estimate which
ground state has the lowest energy one can compare the umklapp gap with that due
to the twist instability. For $K_{c+}\ll 1$ the former scale as $\Delta_u \sim
\frac{v_F a}{R^2}$ whereas the latter as $\Delta\sim \frac{v_Fa^2}{R^3}$.
Therefore in wide tubes the umklapp processes win. The outcome of the competition
in narrow tubes depends on the values of the umklapp coupling constants.
Theoretical estimates~\cite{Odintsov1999} for the gaps range from tens to
hundreds of Kelvins for $(5,5)$ tubes. Experimentally, armchair carbon nanotubes
with $R\sim 0.8 \, nm$ show no sign of the gap down to $T= 4K$~\cite{Dai2003}.
Using $1/R^2$ scaling we infer that the umklapp gaps in $(5,5)$ armchair tubes
should be below our estimates of $70 \, K$ for the twist gap. These estimates
suggest that twist instability may occur in freely suspended $(5,5)$ tubes. It
could be observed by measuring the twist angle in the torsion resonators similar
to those studied in Ref.~\onlinecite{Washburn}.

We would like to thank David Cobden and Dam Son for useful
discussions. This work was supported by DOE grants DE-FG02-07ER46452
(W.C. and A.V.A.) and DE-AC02 -98 CH 10886 (A.M.T.), by the the BNL
LDRD grant 08-002 (A.M.T.), and the US-Israel Binational Science
Foundation grant No. 2004162 (D. O.).

\end{document}